\documentclass{iaus}
\usepackage{graphicx}

\newcommand{\simless}{\mathbin{\lower 3pt\hbox
{$\rlap{\raise 5pt\hbox{$\char'074$}}\mathchar"7218$}}} 
\newcommand{\simgreat}{\mathbin{\lower 3pt\hbox
{$\rlap{\raise 5pt\hbox{$\char'076$}}\mathchar"7218$}}} 

\title[The PMS of A stars] 
{The Pre--Main-Sequence of A-type stars}

\author[Marconi \& Palla]   
{Marcella Marconi$^1$ and Francesco Palla$^2$}

\affiliation{$^1$INAF-Osservatorio Astronomico di Capodimonte, Via Moiariello
  16, 80131 Napoli, Italy\break email: marcella@na.astro.it\\[\affilskip]
$^2$INAF-Osservatorio Astrofisico di Arcetri, Largo E. Fermi 5, 50125 Firenze,
  Italy\break email: palla@arcetri.astro.it}

\pubyear{2004}
\volume{224}  
\pagerange{119--126}
\date{?? and in revised form ??}
\jname{The A-Star Puzzle}
\editors{J.\,Zverko, W.W.\,Weiss, J.\,\v{Z}i\v{z}\v{n}ovsk\'{y}, \& S.J.\,Adelman, eds.}
\begin{document}

\maketitle

\begin{abstract}
Young A-type stars in the pre--main-sequence (PMS) evolutionary phase are
particularly interesting objects since they cover the mass range ($\sim$1.5-4
M$_\odot$) which is most sensitive to the internal conditions inherited
during the protostellar phase. In particular, they undergo a process of
thermal relaxation from which they emerge as fully radiative objects
contracting towards the main sequence.  A-type stars also show intense
surface activity (including winds, accretion, pulsations) whose origin is
still not completely understood, and infrared excesses related to the
presence of circumstellar disks and envelopes. Disks display significant
evolution in the dust properties, likely signalling the occurrence of
protoplanetary growth.  Finally, A-type stars are generally found in multiple
systems and small aggregates of lower mass companions.

\end{abstract}

\firstsection 
\section{Introduction}

Young A-type stars do not resemble their mature siblings.  As a class, they
are known  as Herbig Ae stars (Herbig 1960) because of the presence of
optical emission lines in their spectra, their association with nebulosities,
and conspicuous infrared excess in the spectral energy distributions.  Their
early evolution is marked by the occurrence of a variety of phenomena that
disappear in the course of time.  When they emerge from the protostellar
phase, the internal structure is highly unrelaxed and must undergo a global
readjustment. Unlike low-mass protostars which are fully convective,
intermediate mass protostars (in the mass range $\sim$1.5 and 4 M$_\odot$ )
have developed a radiatively stable core and an outer convective region where
deuterium burns in a shell (Palla \& Stahler 1990). As a result, stars in
this mass range begin with a modest surface luminosity and undergo thermal
relaxation in which the central regions contract while transferring heat to
the expanding external regions. Then, in a short time, the star acquires its
full luminosity and begins contracting towards the main sequence (Palla \&
Stahler 1993).

Another sign of distinction of young A stars is their surface activity.
Winds are relatively common at rates $\sim
10^{-8}-10^{-7}$~M$_\odot$~yr$^{-1}$, but their origin is not understood
(Corcoran \& Ray 1997; Catala \& Boehm, T. 1994). Evidence for accretion at
similar rates is also available from ultraviolet lines and redshifted
Lyman~$\alpha$ lines (Deleuil et al. 2004). Equally unexplained is the X-ray
emission with luminosities intermediate between those in T Tauri and massive
stars (Hamaguchi et al. 2001). Also, young A stars rotate more rapidly than
lower mass objects, but still significantly below breakup and below the
values observed in main sequence stars of the same spectral type (Boehm \&
Catala 1995). Thus, they should spin up considerably during (PMS)
contraction, assuming conservation of angular momentum. Additionally, their
outer layers are subject to $\delta$~Scuti-like pulsational instability,
albeit for a limited amount of time (Marconi \& Palla 1998).  Finally, and
quite importantly, circumstellar disks are almost invariably found around
them with characteristics that indicate significant evolution during the PMS
phase.

Once they reach the main sequence, A-type stars continue to show interesting
properties, but most of the excitement is gone. The major changes take place
in a very short time: from $\sim$17 Myr for a 1.5 M$_\odot$ star to
$\sim$1~Myr for a 3.5 M$_\odot$~object. In the following, we shall provide an
overview of the main properties and the many puzzles that characterize the
brief, but intense early life of A stars.

\section{Evolutionary properties}

As stated above, the initial phases of PMS evolution are marked by the
persistence of the conditions inherited during protostellar accretion.
Intermediate mass stars are the most affected ones by the prior
history and their evolution departs significantly from the standard results
established in the works of Hayashi, Iben and collaborators in the 1960s. A
more modern description of PMS evolution in the Hertzprung-Russell
(HR) diagram from the birthline to the ZAMS is shown in figure 1 (\cite[Palla
\& Stahler 1999]{PS99}). The birthline is the locus where stars first appear
in the HR diagram after the main accretion phase and the results displayed
here have been obtained assuming a protostellar mass accretion rate of
$\dot M_{\rm acc}=10^{-5}$~M$_\odot$~yr$^{-1}$, typical of the observed
values in dense molecular cores undergoing dynamical collapse (e.g., Lee
et al. 2004).
Although quantitatively the initial conditions change with $\dot M_{\rm acc}$, 
the qualitative behavior of protostars is not changed substantially (see
a discussion in Palla 2002).

 \begin{figure}
\begin{center}
\includegraphics[height=3in,width=3in,angle=0]{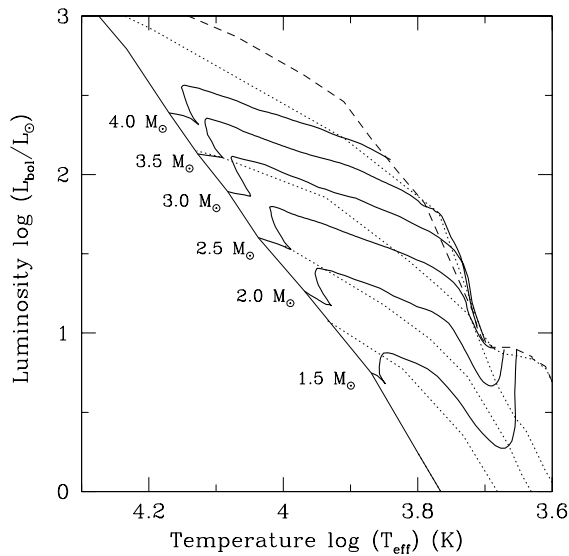}
  \caption{HR diagram of stars in the mass range 1.5--4 M$_\odot$. The 
  evolutionary tracks start at the birthline (dashed line) and end
  on the ZAMS (thin solid line). Selected isochrones (dotted lines) are
  for 0.1, 1, 3, and 10 Myr (from top to bottom). (Adapted from Palla \&
  Stahler 1999)}
\label{fig:tracce}
\end{center}
\end{figure}

The main feature of the new tracks of figure 1 is the complete
absence of a fully convective phase. Starting at about 1.5 M$_\odot$, PMS
stars have almost fully depleted their initial deuterium reservoir and, owing
to their small radius ($\sim$4 R$_\odot$), do not have any energy sources
(nuclear or cooling from the surface) to maintain convection. Thus, their
evolution starts near the bottom of the Hayashi phase and quickly
joins the radiative tracks of homologous contraction.

Stars in the range 2.5--3.5 M$_\odot$ begin their evolution at low luminosity
and move up to join the radiative track. In fact, the tracks loop behind the
birthline and then recross it before pursuing the horizontal paths.  These
stars undergo nonhomologous contraction and thermal relaxation during which
the radius, luminosity, and effective temperature all increase, while the
outer convection disappears. Note also the crowding of the tracks that all
start from about the same position before moving up vertically. Thus,
assigning a correct mass to stars in this part of the HR diagram is tricky.

Finally, stars of mass $\simgreat$4 M$_\odot$ appear immediately on the
radiative track and begin to contract homologously under their own gravity.
As the star contracts, the average interior luminosity rises as
$R_\ast^{-1/2}$, as is evident from the slope of the tracks.

Some important properties of the evolution, both at the beginning and end of
the PMS phase, are given in Table 1. The first two rows list the stellar
radii on the birthline and on the ZAMS. Note the similarity of the values of
R$_{\rm init}$ that explains the crowding of the tracks in figure 1. The table
also lists $\Delta$M$_{con}$, the mass in the convection zone. For the ZAMS
models, this convection is located in the center and is due to the strongly
temperature-sensitive CN burning. Finally, t$_{\rm rad}$ is the time (in
units of $10^6$ yr) when the star becomes fully radiative, measured relative
to the star's appearance on the birthline. Note the dramatic reduction of
t$_{\rm rad}$ for stars more massive than $\sim$3 M$_\odot$. The last entry,
t$_{\rm  ZAMS}$, gives the total duration of the PMS phase.

\begin{table}\def~{\hphantom{0}}
  \begin{center}
  \caption{Evolution of Intermediate-Mass PMS stars}
  \label{tab:kd}
  \begin{tabular}{lcccccc}\hline
Property & 1.5 M$_\odot$ & 2.0 M$_\odot$ & 2.5 M$_\odot$ & 3.0 M$_\odot$ & 
      3.5 M$_\odot$  & 4.0 M$_\odot$\\\hline
 R$_{\rm init}$ (R$_\odot$) & 5.0 & 4.4 & 4.1 & 4.0 & 4.2 & 7.8 \\      
 R$_{\rm ZAMS}$ (R$_\odot$) & 1.7 & 1.8 & 2.1 & 2.3 & 2.5 & 2.7 \\
  & & & & & & \\
 $\Delta$M$_{con}^{\rm init}$ (M$_\odot$) & 1.5 & 2.0 & 1.8 & 1.1 & 0.7 & ... \\
 $\Delta$M$_{con}^{\rm ZAMS}$ (M$_\odot$) & 0.2 & 0.3 & 0.5 & 0.6 & 0.7 & 0.9 \\
  & & & & & & \\
 t$_{\rm rad}$ (Myr)   & 9.4 & 3.4 & 1.3 & 0.4 & 0.07 & ... \\
 t$_{\rm  ZAMS}$ (Myr) & 17  & 8.4 & 3.9 & 2.0 & 1.3  & 0.8 \\
\hline
  \end{tabular}
 \end{center}
\end{table}

\section{Circumstellar disks around A stars}

Once the main phase of accretion is completed, the stellar core emerges as an
optically visible star along the birthline.  The circumstellar matter
around the star, partly distributed in a disk and the rest in an
extended envelope, still emits copiously at infrared and millimeter
wavelengths. The spectral energy distributions of Herbig Ae stars resemble
those observed in classical T Tauri stars (CTTSs), thus suggesting
that the dust responsible for the thermal emission has similar properties and
geometrical distribution (e.g., Hillenbrand et al.  1992). However, there are
important differences.  Unlike CTTSs, a significant fraction of the
luminosity is emitted at short wavelengths with a prominent peak at 2--3
$\mu$m (Meeus et al. 2001). These properties cannot be explained in terms of
standard disk models where the dust is heated by viscosity and stellar
radiation, and is distributed in optically thick and geometrically thin or
flared disks. Thus, some basic modifications of the disk structure are
required. For example, Dullemond et al. (2001) have suggested that the innermost
regions where the dust sublimates (few tenths of AU) are distributed in an
optically thick puffed-up rim that shadows the optically thin dust at larger
distances (which remains cool), thus explaining the strong NIR emission. 

The predictions of these models and the geometrical shape of the disks can
now be directly probed by means of near- and mid-IR interferometry and
adaptive optics on large telescopes that can resolve regions $\sim$0.1 to a
few AU in size at the typical distances of 0.3--1 kpc. The initial results
indicate that indeed the data are best reproduced by flared passive disks
with puffed-up inner rims (Eisner et al. 2004; Leinert et al. 2004).  A
particularly striking example is the circumstellar disk around AB Aurigae
shown in figure~2   This A0 star is one of the closest (d=144 pc) and best
studied Herbig Ae objects, with mass $\sim$2.5~M$_\odot$
and age $\sim$4~Myr. Millimeter observations of $^{13}$CO gas have revealed
the presence of a rotating disk $\sim$450 AU in radius and an estimated mass
$\sim$0.02~M$_\odot$ (Mannings \& Sargent 1998). The disk is immersed in an
extended envelope ($>$1000 AU) visible in scattered light. The image in
figure~2 shows that the extended emission has a double spiral structure almost
the same size of the CO disk and is possibly associated with the latter
rather than the envelope (Fukagawa et al. 2004).

 \begin{figure}
\begin{center}
\includegraphics[height=3in,width=3in,angle=0]{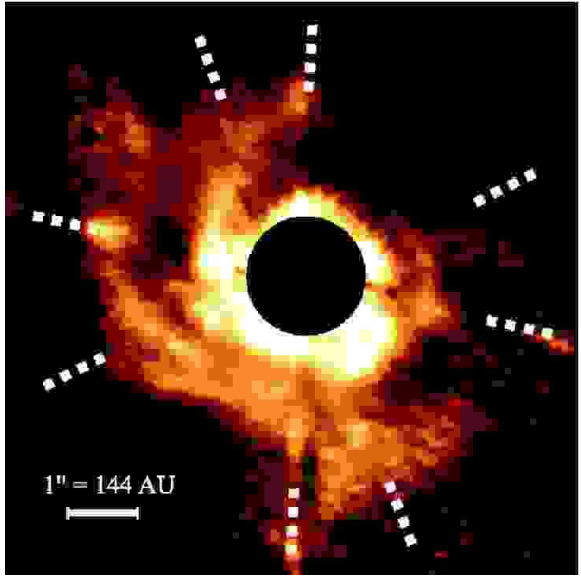}
  \caption{Disk around AB Aur. H-band image at resolution 0.1 arcsec obtained
  wit a coronagraphic imager on Subaru (Fukagawa et al. 2004).}
\label{fig:abaur}
\end{center}
\end{figure}

That disks are common around Ae stars is directly seen at millimeter
wavelengths where spatially resolved images reveal disk features at scales of
few hundred AU in CO emission (Dutrey 2004).  Interestingly, while keplerian
disks are found between 75 and 100\% of Ae stars, the percentage drops
dramatically for the more massive Herbig Be stars. This is shown in the upper
panel of figure~3 that covers two orders of magnitudes in stellar mass (from
CTTSs to B0 type stars). A likely explanation of this trend is related to the
rapid dispersal of the disks around more massive and luminous stars which are
subject to strong radiation UV fields and more powerful outflows.  Finally,
the lower panel of figure~3 reveals that although the disk mass increases with
stellar mass, the ratio of the disk-to-star mass remains basically  constant
in the range 0.2--3 M$_\odot$, implying that all PMS stars have low-mass
disks.

As a last property of disks, let us point out that A-type stars provide the
strongest evidence for disk evolution during the PMS phase and on the MS.
The time evolution of the disk mass is displayed in the right panel of figure~3 covering more
than 4 decades in stellar ages. For ages less than $\sim$10 Myr, the disk
mass is about constant. Then, in objects on the ZAMS, such as HR~4796 and
$\beta$~Pic, the disks are significantly reduced in mass and almost
completely devoid of gas (so-called {\it debris disks}). At this stage, there
is also evidence for grain growth in which dust gradually agglomerates into
larger, more crystalline structures that should favor planetesimal
condensation.  The dust grains are subject to the Poyting-Robertson drag from
stellar photons which causes them to spiral inward. Others must be
resupplied, presumably through the mutual collisions of larger orbiting
bodies.  In fully evolved stars ($\alpha$~Lyr and $\alpha$~PsA,
$\simgreat$200~Myr) the dust becomes undetectable in scattered light and visible
only in the far-IR and its actual luminosity has declined to only $\sim
10^{-5}$~L$_\ast$.

\begin{figure}
\hbox{
\vbox{
\includegraphics[height=3in,width=2.7in,angle=0]{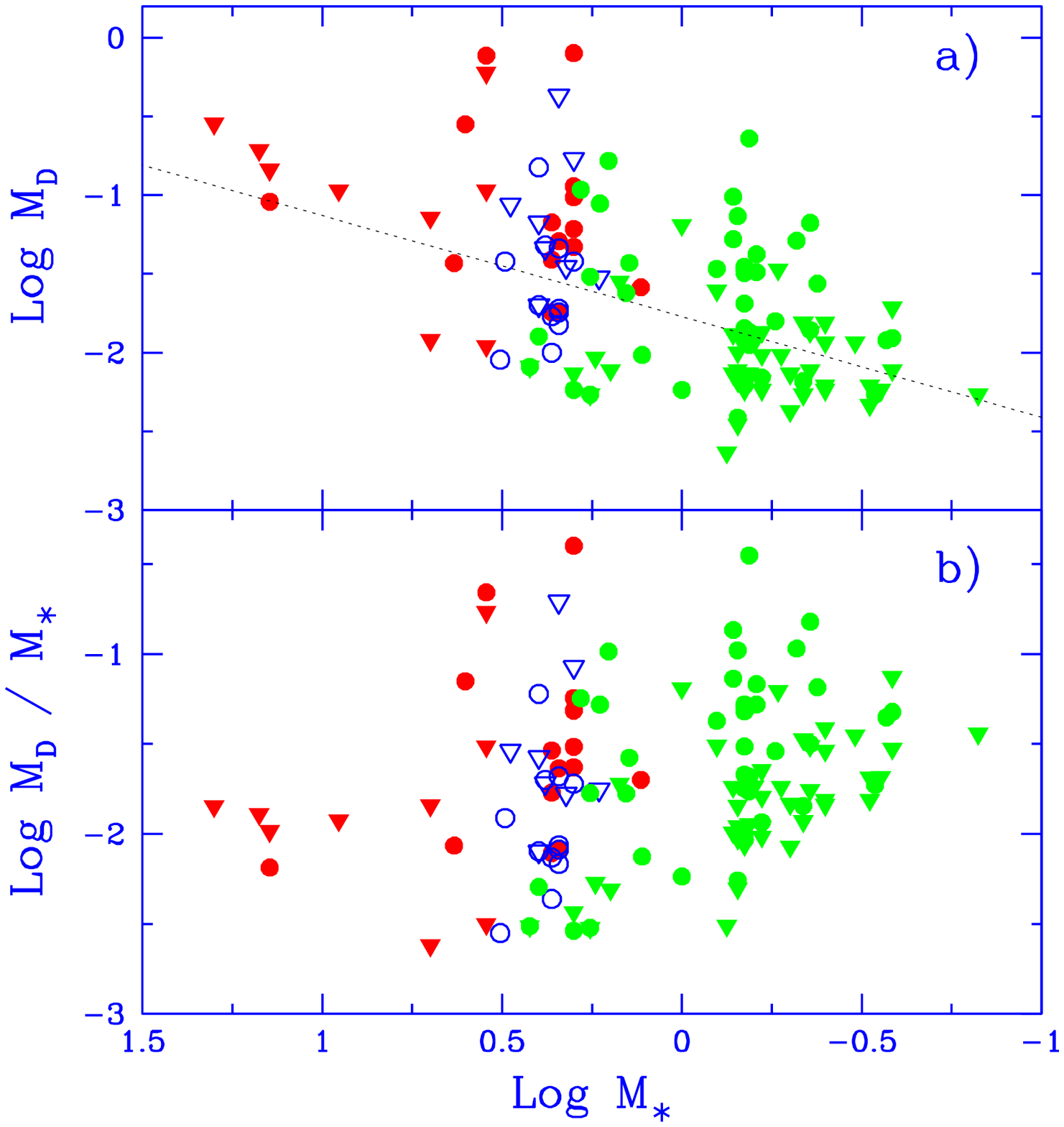}
\includegraphics[height=3in,width=2.7in,angle=0]{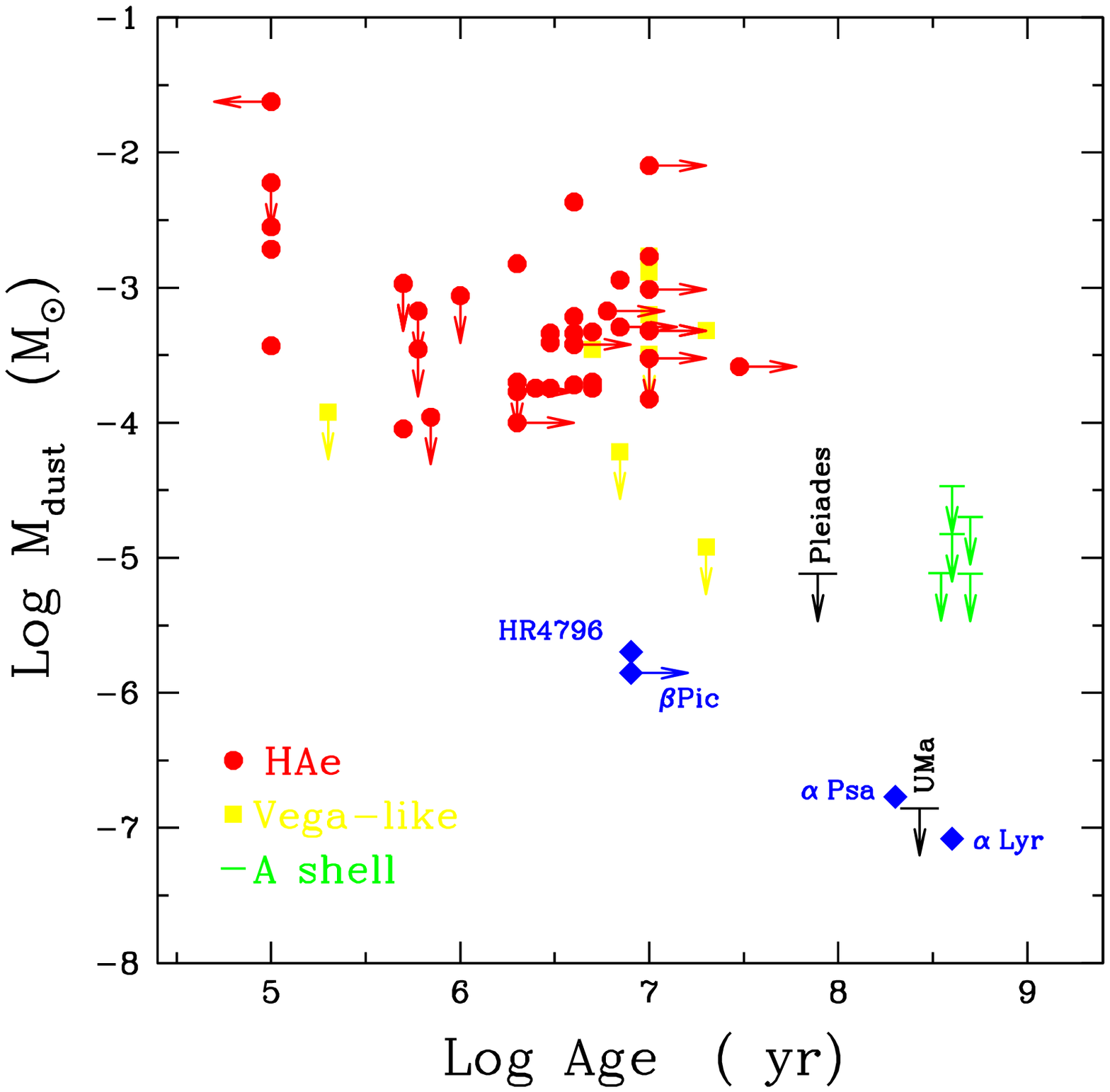}
}}
 \caption{Left: Properties of disks around low- and intermediate-mass stars.
  {\it Upper panel}: disk mass as a function of the stellar
  mass as inferred from mm-interferometric observations. Filled and empty
  circles are detections, while triangles are upper limits.
   {\it Lower panel}: ratio of the disk-to-star mass {\it vs.} stellar mass
   (Natta 2003). Right: Evolution of the disk mass with time for A-type stars
 (Natta et al. 2000).}
\end{figure}

%

\section{Interaction with the environment}

Most Herbig Ae stars are found close to or embedded within their natal gas
clouds. Studies of the distribution of the $^{13}$CO emission reveal the
pattern shown in figure~4 where the youngest stars are still immersed in dense
cores (top) while the older ones are found in cavities created during PMS
evolution (Fuente et al. 2002). The physical process responsible for the
removal of the dense gas is likely to be related to the activity of bipolar
outflows during the protostellar phase. Bipolar outflows sweep out matter
along the poles and creat a biconical cavity. Material in the envelope
accretes onto a circumstellar disk that feeds the growing star. In this way,
the cloud evolves toward a more centrally peaked morphology while a
significant fraction of the core material is dispersed. Assuming a typical
accretion rate of $10^{-5}$~M$_\odot$~yr$^{-1}$, an A-type star (2-4
M$_\odot$) will be formed in a few$\times$0.1~Myr: by this time, about 90\%
of the dense core results dispersed. From a few$\times$0.1~Myr to
$\simeq$1~Myr, the star begins the PMS contraction and only a small amount of
circumstellar matter is removed because (a) the outflow activity fades
rapidly in time, and (b) the stellar surface is still too cold to generate UV
photons that can photodissociate the surrounding gas. Finally, from
$\geq$1~Myr to $\sim$10~Myr the star reaches the ZAMS and completes the
creation of the cavity thanks to the UV radiation, albeit at a slow rate due
to the low effective temperatures. This evolutionary sequence highlights once
more the transitional character of A-type stars: the late-type A stars behave
like T Tauri stars and their associated weak winds, whereas the early-type A
stars begin to show the phenomenology (both radiative and mechanical)
associated with the more luminous Be-type objects.

\begin{figure}
 \includegraphics[height=4in,width=4in,angle=0]{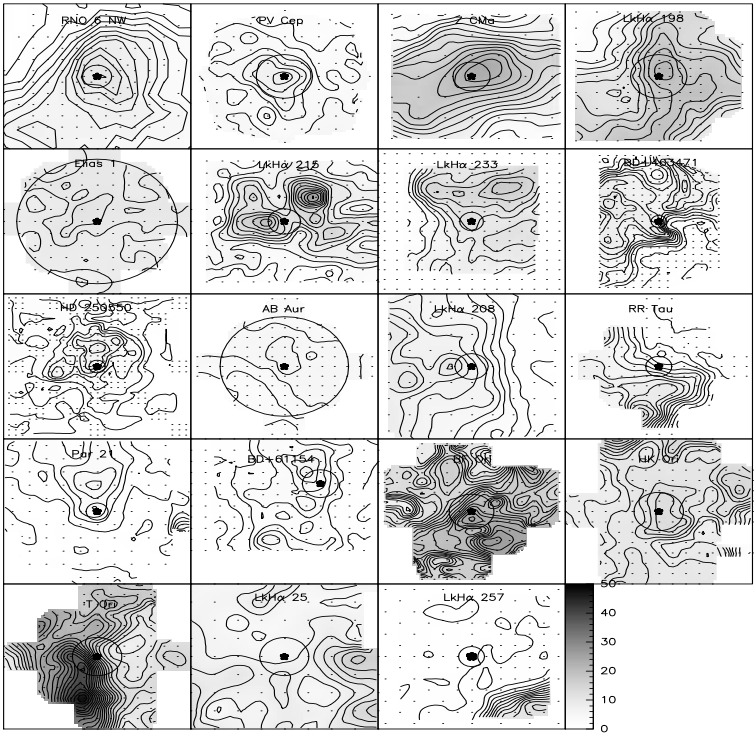}
 \caption{The birth sites of Ae stars seen in $^{13}$CO emission. The
 size of the box varies from source to source. In each panel, a circle of
 0.08 pc in radius is drawn around the central star (Fuente et 
 al. 2002).}
 \label{fig:CO}
\end{figure}

In addition to the properties of the gas, it is interesting to consider the
issue of stellar companionship. It is well known that most stars are found in
binary or multiple systems and Herbig Ae stars are no exception. The binary
frequency of A0-A9 stars with semi-major axis less than 2000 AU
(corresponding to log P=5-7 days) is 42\%, or 50-60\% when corrected
for incompleteness (Bouvier \& Corporon 2001). For comparison, the binary
frequency of field G dwarfs in the same period interval is only 20\%. Thus, binary Herbig Ae stars exceed binary G-type stars by a factor of 2 and show an higher percentage than binary T Tauri stars.
This difference certainly has some bearing on the formation process. In
particular, it is possible that Herbig Ae stars form in more crowded
environments than lower mass stars and that dynamical interactions lead to
capture of neighboring stars. Indeed, models of the evolution of small
clusters predict an increase of the binary frequency with the mass of the
primary. Interestingly, Ae stars tend to be found in aggregates with density
$\simless 10^2$~stars~pc$^{-3}$ (Testi et al.  1999), higher than the
isolated regime of T Tauri stars ($\simless$10~stars~pc$^{-3}$), but 
lower than the clusters associated with massive stars ($\simgreat
10^3$~stars~pc$^{-3}$).

%

\section{Variability}

Young A-type stars show evidence of dusty environment, intense stellar
activity and strong stellar winds. As a result they are photometrically,
spectroscopically and polarimetrically  variable on very different time
scales and wavelength domains (e.g., Herbst \& Shevchenko 1999).  The typical
long-term variability due to obscuration by circumstellar dust  is called
{\it UX Ori} variability from the name of the prototype object (see figure 5).
The periodicity has a time scale on the order of $\sim$1~yr and can be as large as 3
magnitudes in the V band. After a few days at minimum, the star resumes the
normal brightness in several weeks. During the approach to minimum, the color
becomes bluer and the polarization increases dramatically. Both effects
suggest that the origin of the dimming is due to scattering by circumstellar
dust distributed in optically thick clouds that partially occult the star
(e.g. Grinin et al. 1991). The other type of variability takes place on
shorter time scales (typically, few hours) and involves much smaller
variations (few thousandths to few hundredths of mags). Since the variability
is associated with the $\kappa$ mechanism, its study allows to peer into the
the intrinsic properties (and through asteroseismology also the internal structure) of A-type stars, as we will describe below.

\begin{figure}
\begin{center}
 \includegraphics[height=3in,width=4in]{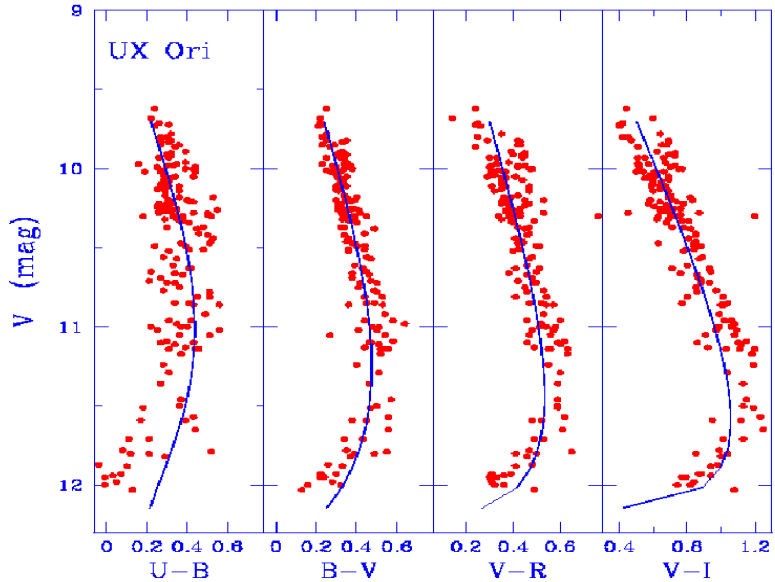}
 \caption{Color-magnitude diagrams of UX Ori showing the blueing effect at
 minimum(Rostopchina et al. 1999).}
 \label{fig:uxori}
\end{center}
\end{figure}

\subsection{The $\delta$ Scuti type pulsation}
During the contraction phase toward the Main Sequence, intermediate-mass
stars cross the pulsation instability strip of more evolved variables,
suggesting that, in spite of the relatively short time spent in the strip
($\sim10^5$-$10^6$ years), at least part of the observed activity could be
due to intrinsic variability.

 \begin{figure}
 \begin{center}
\includegraphics[height=3in,width=3in,angle=0]{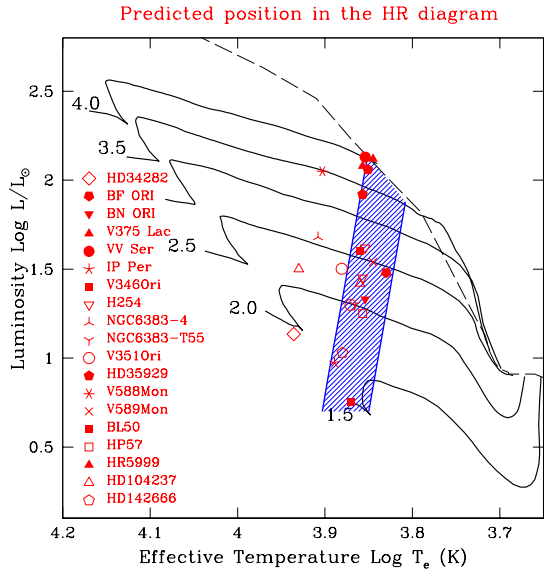}
  \caption{The position of PMS $\delta$ Scuti stars in the HR diagram as 
  predicted on the basis of the comparison between the observed periodicities 
  and linear nonadiabatic radial pulsation models. The dashed region is the 
  theoretical instability strip for the first three radial modes (Marconi \& 
  Palla 1998), that is the region between the second overtone blue edge and the fundamental red edge.}
  \label{fig:strip}
\end{center}
\end{figure}

The first observational evidence for such variability is due to Breger (1972)
who detected  $\delta$ Scuti-like pulsations in two Herbig stars of the young
cluster NGC~2264, namely V588~Mon and V589~Mon. The issue was reconsidered
more than 20 years later by  Kurtz \& Marang (1995) and Donati et al.  (1997)
who found $\delta$ Scuti-like variability in  the Herbig stars HR~5999 and
HD~104237.  Since then there has been a renewed interest in the study of
these young pulsators, both from the observational and theoretical point of
view.  For the latter, using convective nonlinear models, Marconi \& Palla
(1998) computed  the first theoretical instability strip for PMS $\delta$
Scuti stars (see figure 6). They also identified a list of candidates with
spectral types in the range of the predicted instability region.  This
theoretical investigation stimulated new observational programs carried out
by various groups with the result that the current number of known or
suspected candidates amounts to about 20 stars.

A census of known or suspected PMS $\delta$ Scuti stars is reported in Table
2, whereas the position of these pulsators in the HR diagram, as resulting
from the comparison between the observed and predicted pulsation frequencies,
is shown in figure 6.  We notice the good agreement between the predicted
instability strip and the location of observed pulsators. The only deviating
objects are hotter than the theoretical second overtone blue edge and
are indeed predicted to pulsate in higher overtones, whereas no
pulsating object is found to the right of the theoretical red edge.
The main limitations of this approach are: 1) the uncertainties still
affecting many of the observed frequencies, due to poor data quality
and/or the aliasing problem; 2) the difficulty to discriminate between
PMS and post-MS evolutionary phases on the basis of radial models, in
particular for pulsators that are predicted to be located close to the
MS; 3) the fact that the likely presence of nonradial modes is not
taken into account.

Concerning the first point, significant improvements can be obtained by means
of multisite campaigns and will certainly be obtained with future satellite
missions (e.g.  EDDINGTON and COROT). As for the last two issues, it is clear
that both radial and nonradial models should be computed in order to better
understand the intrinsic properties of the rather unexplored class of young
variable stars.  In order to cope with this problem, we have started a
project to apply the adiabatic nonradial code by Christensen-Dalsgaard (available on the web page http://astro.phys.au.dk/~jcd/adipack.n)
to  PMS evolutionary models. Preliminary results seem to suggest the 
coexistence of radial and nonradial frequencies at least in the young 
pulsators V351 Ori and IP Per.

\begin{center}
\begin{table}
\tiny
\caption[]{Pulsation properties of known or suspected PMS $\delta$ Scuti stars.\label{tab1}}
\begin{tabular}{lcccccccc}
\hline
Name      &     F1              & F2                   & F3                &
F4              & $\Delta$V &  V     & Sp.Ty. & Ref. \\
          &     (c/d)           & (c/d)                & (c/d)             &
(c/d)           &   (mag)   &  (mag)     &    &  \\
\hline
V588 Mon  & 7.1865$\pm$±0.0006   & ?                    & ?                 &                &           0.04   &  9.7   & A7   & 1  \\
V589 Mon  & 7.4385$\pm$±0.0006   & ?                    & ?                 &                 &           0.04   &  10.3  & F2    & 1 \\
HR5999    & 4.812$\pm$±0.010     &                     &                   &
                &            0.02   &  7.0   & A7    & 2 \\
HD104237  & 33$\pm$±0.2                        &                   &
                 &       &    0.02   &  6.6   & A7    & 3 \\
HD35929   & 5.10$\pm$±0.13       &                      &                   &                 &           0.02   &  8.1   & A5    & 4 \\
V351 Ori  & 15.687$\pm$±0.002    & 13.337 $\pm$±0.002    & 16.868$\pm$±0.002 & 11.780$\pm$±0.002  &          0.1   &  8.9   & A7    & 5 \\
BL 50     & 13.9175$\pm$±0.0005 & 9.8878$\pm$±0.0009     &                   &            &          0.02   &  14.5  & --    & 6 \\
HP 57     & 12.72557$\pm$±0.0002& 15.52437$\pm$±0.0003 &                     &     &            0.03   &  14.6  &       --   & 6  \\
HD142666  & 21.43$\pm$±3         &                      &                   &         &          0.01   &  8.8   & A8     & 7 \\
V346 Ori  & 35.3$\pm$±2.3        & 22.6$\pm$±2.7         & 45.5 $\pm$2.5
   & 18.3$\pm$±2.5              &   0.015   &  10.1  & A5    & 8 \\
H254      & 7.406$\pm$±0.008     &                      &                   &     &           0.02   &  10.6  & F0    & 9 \\
NGC6383-4 & 14.376              & 19.436               & 13.766            & 8.295              &   0.014   &  12.61 & A7    & 10  \\
NGC6383-T55 &   19.024            &                &            &               &  0.002    & 20.9 &  -- &  10 \\
IP Per &    22.887            &  34.599              &  30.449           &  48.227             & 0.004     & 10.35  & A7    &  11 \\  
V375 Lac &    5.20$\pm$0.14           &  2.40$\pm$0.14               &   9.90$\pm$0.14      &               &  0.004    & 13.56  & A7   & 12  \\
VV Ser &  5.15$\pm$0.01              &   8.61$\pm$0.01             &       4.46$\pm$0.01   &               &  0.005    & 11.87  &  A2(?)  & 12  \\
BN Ori &   10-12.7             &                &             &               &  0.002    & 9.67  & F2 (?)   &  12 \\
BF Ori &    5.7$\pm$0.3            &                &             &               &   0.006   & 10.41  & A5 (?)   & 12 \\
HD34282 &      79.5$\pm$0.06          &   71.3$\pm$0.06             &             &               &  0.011    & 9.873   &  A0-A3   & 13  \\
IC4996-37 &    31.87            &                &             &               &  0.005    & 15.302  &  A5    &  14 \\
IC4996-40 &    42.89            &                &             &               &  0.008    & 15.028   &  A4   &  14 \\
\hline
\end{tabular}

\tiny
Sources: (1) Breger (1972), Pe\~na et al. (2002); 
(2) Kurtz \& Marang (1995), Kurtz \& Catala (2001); 
(3) Donati et al. (1997), Kurtz \& Muller (1999); 
(4) Marconi et al. (2000); 
(5) Ripepi et al. (2003); 
(6) Pigulski et al. (2000); 
(7) Kurtz \& M\"uller (2001); 
(8) Pinheiro et al. (2003); 
(9) Ripepi et al. (2002); 
(10) Zwintz et al. (2004); 
(11) Ripepi et al. 2004;
(12) Bernabei et al. 2004;
(13) Amado et al. 2004;
(14) Zwintz 2004 private communication
\end{table}
\end{center}

\section{Concluding remarks}

Although PMS A-type stars cover a small mass interval ($\sim$1.5--4
M$_\odot$), they represent an interesting laboratory for the study of a
variety of physical processes, many of which are still poorly understood.  As
a class, they share many of the characteristic properties of both the lower
mass T Tauri stars and of the more massive Herbig Be objects, but depart from
them in significant ways.  Under the influence of protostellar evolution,
A-type stars begin their PMS evolution as thermally unrelaxed structures that
undergo a global readjustment from partially convective to fully radiative
interiors in a short time scale.  Circumstellar disks are much more frequent
than in Herbig Be stars, and at least as common as in T Tauri stars. However,
the more intense radiation field from the hotter stars makes standard viscous
disk models insufficient to explain the observed infrared emission, requiring
important modifications in the disk structure and geometry.  Unlike Herbig Be
stars, PMS A-type stars are generally found in relative isolation or in
aggregates containing $\sim$10-100 lower mass members.  Finally, their youth
is marked by an intense surface activity, which includes  winds, accretion,
fast rotation, X-ray emission, variable obscuration from circumstellar dust
and also intrinsic $\delta$ Scuti type pulsation. The latter, with its
asteroseismological implications, can offer a unique tool to study their
internal structure, to test evolutionary models, and to obtain independent
estimates of the stellar mass, the fundamental parameter governing stellar
evolution.

\begin{discussion}

\discuss{Budovicov\'a}{Why Be stars do not have a gaseous disk? It is one of
their basic properties. Why is it different from Ae stars (having a gas disk)
?}

\discuss{Marconi}{I first remind that we are talking of PMS A-type and B-type
stars. The lower frequency of disks around the latter may be due to rapid
destruction due to photoevaporation and/or winds. However, there are recent
detections of circumstellar disks around Herbig Be stars (e.g., Fuente et
al.  2003 ApJ 598 39).}

\discuss{Piskunov}{T\,Tauri stars come in two flavours: classical and
weak-line. Do we see a similar phenomenon in pre-MS A stars?}

\discuss{Marconi}{No, the same distinction does not hold for Herbig Ae (or
Be) stars. Although many young clusters contain a large population of
intermediate mass PMS stars, only a small minority of these objects qualify
as being genuine Herbig-type stars. As George Herbig suggested, this may be
indication that either the Ae/Be phenomenon is a temporary one, or that some
stars in that mass range do not show it at all. In this sense there
might be two classes of A-type PMS stars.}

\end{discussion}

\end{document}